\documentstyle[aps,twocolumn,epsf]{revtex}

\begin{document}
\title{Quantum communication protocols using the vacuum}
\author{S.J. van Enk, and Terry Rudolph \\
Bell Labs, Lucent Technologies,\\
600-700 Mountain Ave,
Murray Hill NJ 07974}
\maketitle
\begin{abstract}
We speculate what quantum information protocols can be
implemented between two accelerating observers using the vacuum. 
Whether it is in principle possible or not 
to implement a protocol 
depends on whether the aim is to end up with classical 
information or quantum information.
Thus, unconditionally secure coin flipping
seems possible but not teleportation. 
\end{abstract}

\medskip
\section{Introduction}
Entanglement is in the eyes of the beholder. Unitary
transformations that affect the definition of one's systems
change, in general, the amount of entanglement in a given state.
For instance, the Lorentz transformation of the spin degrees of
freedom of a particle depends on its momentum. Thus the
entanglement between the spin and momentum of a particle is not a
Lorentz-invariant concept \cite{scudo}. One observer may see a
product state, while another believes there is entanglement.
Similar discussions of the Lorentz invariance or the lack thereof 
of entanglement can be found in Refs.~\cite{alsing,gingrich}.
 For another example, take a state containing a single circularly
polarized photon. When written in terms of linear polarization,
this state becomes $(|0\rangle|1\rangle +i |1\rangle |0\rangle)/\sqrt{2}$,
which seems to be maximally entangled (for similar but more 
complicated and less boring cases see
\cite{em}). In both types of examples, however, the apparent
entanglement is always {\em local}. As such, it cannot be used for
any nontrivial quantum communication protocols.

Here we discuss a well-known phenomenon that does produce {\em
nonlocal} entanglement, the Unruh effect \cite{unruh,audretsch,pringle}. 
It involves just the
vacuum, and the unitary transformations arise when one describes
accelerating observers. We investigate which, if any, quantum
communication protocols could be implemented using this resource.
We consider protocols involving two accelerated observers, in contrast to
Ref.~\cite{alsing2}, which considers quantum
teleportation involving one inertial 
and one accelerating observer.

\section{The Unruh effect}
Suppose Alice is accelerating  at a uniform acceleration $a$. As
is well known \cite{unruh}, she will perceive the Minkowski vacuum
state (of, say, the electromagnetic field) as a mixed thermal
state with equivalent temperature $k_BT=\hbar a/(2\pi c)$. Since
the transformation from an inertial to an accelerating frame is
unitary, however, the vacuum should be transformed into a pure
state, not a mixed state. Indeed, the state of the modes inside
Alice's event horizon only appears mixed because it is entangled
with modes that lie outside that horizon. More precisely, each
mode is entangled with one ``mirror'' mode, a mode propagating
along a trajectory that is the mirror image relative to the
appropriate event horizon. For each pair of mirror modes of
frequency $\omega'$ (as measured by the accelerating observers),
the entangled state is in fact a two-mode squeezed state of the
form
\begin{equation}\label{ent}
|\Psi\rangle=\sqrt{1-\mu^2}\sum_n \mu^n |n\rangle |n\rangle,
\end{equation}
where $\mu=\exp(-\pi\omega' c/a)$.
Mirror modes are the appropriate modes for an observer Bob accelerating uniformly
at the same acceleration $a$ but in the opposite direction along a
trajectory that is, again, a mirror image relative to the same event
horizon. That the entanglement and nonlocal correlations present in the state (\ref{ent})
may be real in the sense that they can be measured and perhaps even exploited
has been discussed before \cite{alsing,werner,klyshko,reznik,alsing2}.

The Unruh effect can be understood by considering the
transformation between creation and annihilation operators from
Alice's frame of reference to that of a Minkowski observer, Mork.
The transformation is of the form
\begin{eqnarray}\label{trans}
a'&=&(a-\mu \tilde{a}^{\dagger})/\sqrt{1-\mu^2} \nonumber\\
\tilde{a}'&=&(\tilde{a}-\mu a^{\dagger})/\sqrt{1-\mu^2}.
\end{eqnarray}
where we absorb irrelevant phase factors in the definitions of the
mode operators. Here we use notational conventions that primed
operators and variables correspond to accelerating observers, and
that operators with a tilde correspond to mirror modes. The modes
here are assumed to be localized wave packet modes, as constructed
in \cite{audretsch}. The fact that a creation operator appears in
the transformation of an annihilation operator, distinguishes
(\ref{trans}) from standard unitary transformations of modes
\cite{em}.

\section{Quantum communication protocols}

The questions we consider  now are {\em (i)} what useful quantum
information tasks might Alice and Bob perform with the entangled
state (\ref{ent})? {\em (ii)} how does a Minkowski observer, Mork,
describe their actions and make sense of it? After all, according
to Mork, Alice and Bob share nothing but the vacuum and are
causally disconnected, and so it may seem they should not be able
to perform any interesting protocols. We will consider several
quantum information processing protocols that are known to rely on
entanglement and discuss to what extent they can be implemented by
Alice and Bob.

Concerning question {\em (i)}, there is an important distinction
between two types of quantum communication protocols: those that
terminate with at least one party holding a quantum state, and
those that terminate with all parties holding purely classical
information. In the former case the desired quantum states
typically exist only in the eyes of Alice and Bob, and thus, in
the scenarios considered here, only as long as they keep
accelerating. Clearly this does not allow Alice and Bob to ever
communicate, not even classically. This does restrict at least the
usefulness of the protocol and sometimes it prevents the protocol
from being executed at all. In the latter type of protocols,
however, Alice and Bob are {\em both} free to decelerate after
having performed the required quantum operations (since we
presume, hopefully correctly, that classical information, unlike
quantum information, is not affected by deceleration), and thus
may communicate afterwards. This then may lead, apart from practical
considerations, to useful implementations of certain quantum
protocols. Our primary goal here is to discuss in detail an
example of each type, to demonstrate both the potential and the
limitations of vacuum entanglement for quantum communication
protocols. We also briefly mention various other protocols.

Concerning question {\em (ii)} we note that Alice's and Bob's
local operations  appear nonlocal to Mork, and {\em vice versa}.
The parameter $\mu$, which measures the strength of the Unruh
effect and the amount of nonlocality, written in Mork's
coordinates is equal to (using Ref.~\cite{audretsch})
$\mu=\exp(-\pi^2 D/\lambda)$, where $D$ is the distance between
the mirror trajectories and $\lambda$ the wavelength. (In contrast, recall
that for the accelerating observers $\mu$ does not depend on the distnace between the relevant modes.)
In order to
have any appreciable effect, at the moment Alice and Bob wish to
use their entanglement, they must be within a distance $D\sim
\lambda/\pi^2$ of each other, that is, within the coherence length
of the vacuum fluctuations\cite{klyshko}. This is how Mork can
make some physical sense out of the nonlocal character of Alice's
and Bob's actions and of the fact that, counter to Mork's
expectations, some of their protocols seem to work.

\subsection{Teleportation}
Let  us start with one of the more famous protocols, quantum
teleportation \cite{telep}. Indeed, a two-mode squeezed state
of the same form (\ref{ent}) can be
used for exactly that purpose\cite{akira,sam}. 
We also note that teleportation with the resource 
(\ref{ent}) is briefly discussed in \cite{alsing}. 
A later paper by the same authors considers teleportation involving one inertial and one accelerating observer \cite{alsing2}.

Teleportation is a clear example
where the aim is to end up with a quantum state, 
and where classical communication is necessary.
Thus the prospects for Alice and Bob are bleak. Indeed, 
standard teleportation is not possible, but a weaker variant of it is.
This weaker variant is in essence an example of quantum steering 
\cite{steering}---the process whereby a local choice of measurements by Alice 
can steer Bob's half of an entangled state to any ensemble of his local density operator.

We first describe the experiment from Alice's frame of
reference. Alice uses two modes, one mode $T$ contains the state
she wishes to teleport, the other, $E$ contains half of the
entangled state. Similarly, Bob's corresponding mirror modes are
denoted by $\tilde{T}$ and $\tilde{E}$. 
Let us assume Alice wishes to teleport a  coherent state. In order to prepare
that state, Alice
first cools down the mode $T$, i.e., removes all (Rindler) photons from it.
Subsequently she applies a displacement operation. 
(Cooling 
is not a necessary part of the protocol. Alice could just 
teleport the action of the displacement operation on the thermal state as it is. In that case the description becomes more tedious and less clear.
) 
For
convenience, we assume that Bob, too, cools down his mode $\tilde{T}$
although this is not necessary at all for the teleportation
protocol. The most
convenient way to describe teleportation then is by using the
Wigner function\cite{sam}, as at all times the states of the
modes involved will be Gaussian.
Moreover, it is customary to use Hermitian quadrature variables $X$ and $P$
instead of $a$ and $a^{\dagger}$, defined through
$a=X+iP$, and corresponding eigenvalues $x$ and $p$.
For the two-mode squeezed state
one has (leaving out irrelevant normalization factors)
\begin{eqnarray}
W'_{E,\tilde{E}}\sim
\exp\big(
- \frac{1+\mu}{1-\mu}
 [(x'_{E}-\tilde{x}'_{E})^2+(p'_{E}+\tilde{p}'_{E})^2]\nonumber\\
-\frac{1-\mu}{1+\mu}
[(x'_{E}+\tilde{x}'_{E})^2+(p'_{E}-\tilde{p}'_{E})^2] \big).
\end{eqnarray}
Similarly, a coherent state $|\alpha_0\rangle_{T'}$  is described by
\begin{equation}
W'_{T}\sim \exp(-2(x'_{T}-x_0)^2-2('p_{T}-p_0)^2),
\end{equation}
where $\alpha_0=x_0+ip_0$.

Alice then performs a joint measurement  on her modes $T$ and $E$.
She measures the commuting observables $X'_{E}+X'_{T}$ and
$P'_{E}-P'_{T}$, for example, by homodyne detection
with a strong local oscillator field. After this measurement,
which we assume to have outcomes $X$ and $P$ respectively,
Alice then ascribes the following quantum state to mode $\tilde{E}$ on Bob's
side
\begin{eqnarray}\label{W}
W'_{\tilde{E}}\sim
 \exp \big(-\frac{2-2\mu}{3+\mu}[(x'_{\tilde{E}}+x_0-X)^2
+(p'_{\tilde{E}}+p_0-P)^2]
\nonumber\\
-\frac{2+2\mu}{3-\mu}[(x'_{\tilde{E}}-x_0-X)^2+(p'_{\tilde{E}}-p_0-P)^2    \big)
\end{eqnarray}
When $\mu$ approaches unity,  the state on Bob's side reduces to a
coherent state, but displaced by an amount $\beta=X+iP$.  In the
standard teleportation protocol Alice would send Bob the classical
outcomes $X$ and $P$ and Bob would displace his state by an amount
$-\beta$ to retrieve the state Alice teleported. Clearly, this
step is not possible. According to Bob, his state will always
remain a mixed thermal state, but based on Alice's information she
assigns Bob's system the state (\ref{W}).

It is easy, albeit somewhat tedious, to write down the Wigner
functions Mork observes, by using the inverse transformations of
(\ref{trans}). Clearly,  they will remain Gaussians for Mork as
well. Here, however, we just focus on the local vs. nonlocal
aspects of Alice's actions. Alice and Bob start out with the
vacuum. Then both Alice and Bob cool down their modes $T$ and
$\tilde{T}$. Mork, though, does not view these cooling operations
as local. In fact, he will claim that Alice and Bob create a
two-mode squeezed state between those two modes, of the same form
(\ref{ent}) but with $\mu\rightarrow -\mu$, as follows directly
from the inverse transformation of (\ref{trans}). Alice
subsequently applies the displacement operator to her mode:
according to Mork this is again a nonlocal transformation,
displacing not only mode $T$ but Bob's mode $\tilde{T}$.

Then Alice performs her joint measurement, which according to Mork
is a nonlocal measurement
of the variables
\[\frac{x_{E}+x_{T} +\mu (x_{\tilde{E}}+x_{\tilde{T}}) }{\sqrt{1-\mu^2}},\]
and
\[\frac{p_{E}-p_{T} -\mu (p_{\tilde{E}}-p_{\tilde{T}}) }{\sqrt{1-\mu^2}},\]
with the same outcomes $X$ and $P$, respectively. This will create an entangled
state of all 4 modes involved, according to Mork. This is in contrast to Alice's description,
who believes her 2 modes have been disentangled from Bob's modes.
According to Mork, no teleportation takes place.
\subsection{Secure coin flipping}

Two-party cryptographic protocols involve two antagonistic
parties, Alice and Bob, who wish to complete some information
processing task. In classical information theory it has been
proven that no two-party protocols exist which have ``information
theoretic'' security \cite{wcf,ambainis,coin}. 
In quantum cryptography protocols
\textit{do} exist with various degrees of quantum information
theoretic security, and thus examining these protocols provides a
readily quantifiable way of distinguishing classical from quantum
information theory.

It is standard to assume in two-party  quantum cryptography that
the initial state of systems held by Alice and Bob is separable,
i.e. of the form $\left| 0\right\rangle_A \left|
0\right\rangle_B$. However it is interesting to note that if Alice
and Bob share prior {\em trusted} entangled states then some (otherwise
impossible) arbitrarily secure quantum cryptographic protocols
become possible, while others remain impossible. For example, if
they share a maximally entangled state of two qubits, then an
arbitrarily secure coin flip is trivially possible - the coin flip
outcome is simply the result each party obtains by measuring their
half of the entangled pair in an orthogonal basis. By contrast, as
can be deduced from \cite{coin}, the sharing of a prior trusted
entangled state does \emph{not} give Alice and Bob the ability to
perform an arbitrarily secure bit commitment. Thus there is an
intricate hierarchy of the security obtainable in these protocols
with respect to any initially trusted entanglement resources.

What we are proposing here is that the (Minkowski) vacuum state
$\left| 0\right\rangle_A \left| 0\right\rangle_B $ can also be
considered a ``prior trusted'' state. Since, as discussed above,
this is in fact also a (Rindler) entangled state, we surmise that
it can, in fact, be used to implement a secure coin flip.

We first define the task of coin flipping more precisely:

\begin{quote}
\textit{(Strong) Coin Flipping}: Alice and Bob implement a
protocol, at the end of which each infers the outcome of the
protocol to be one of `$0$'$,$ `$1$' or `fail'. If both are
honest, then they agree on the outcome and find it to be $0$ or
$1$ with equal probability. If party $X$ cheats, while his or her
opponent is honest, then $X$ cannot make the probability of the
opponent finding the outcome $0$ to be greater than $1/2+\epsilon
_{X}^{0}$ and cannot make the probability of the opponent finding
the outcome $1$ to be greater than $1/2+\epsilon _{X}^{1}.$ The
parameters $\epsilon _{A}^{0},\epsilon _{A}^{1},\epsilon
_{B}^{0},\epsilon _{B}^{1},$ which specify the security of the
protocol, must each be strictly less than $1/2.$

The protocol is considered arbitrarily secure if, and only if, the
parameters $\epsilon _{A}^{0},\epsilon _{A}^{1},\epsilon
_{B}^{0},\epsilon _{B}^{1}$ can simultaneously be made to approach
zero.
\end{quote}

Intuitively speaking, such a coin flipping protocol is meant to
result in a random bit outcome, and neither Alice nor Bob should
be able to bias the bit value towards either 0 or 1. It has been
shown by Kitaev, that within the standard quantum communication
paradigm wherein Alice and Bob start with a state of the form
$\left| 0\right\rangle_A \left| 0\right\rangle_B $ and build up an
entangled state via rounds of communication, all coin flipping
protocols satisfy $(1/2+\epsilon _{A}^{b})(1/2+\epsilon
_{B}^{b})\ge 1/2$, $b=0,1$. Thus, arbitrarily secure quantum coin
flipping within this paradigm is impossible. The best known
protocols \cite{ambainis,coin} do not even saturate Kitaev's lower bound.

We should emphasize that in two-party cryptographic protocols
there is a basic presumption that each party feels secure about
their own laboratory.  In fact, it is desirable that this need be
the \emph{only} thing they feel secure about - i.e. we presume
that the parties should not have to feel secure about things
outside their own lab.
We note here explicitly that, for instance, changing 
boundary conditions on the
field in one of the two Rindler wedges does not modify the thermal 
spectrum seen
by an observer in the other wedge, as was pointed out 
by Pringle \cite{pringle}.

The protocol we consider here is as follows:

\begin{quote}
\textit{Unruh based coin flipping:}

From Alice's point of view, 
an instance of the protocol is specified by two points in spacetime
chosen to be located inside Alice's lab. 
At time $t=0$
Alice, who is uniformly accelerating with acceleration $a$, is
instantaneously at rest (see Figure~1) in Mork's reference frame. 
At that moment, Alice turns on a
detector $D_2$ (see Fig.~1), if she records a photon then the outcome of the
coin flip is $1$, otherwise it is $0.$ 
In order for her to trust this result she must have 
checked {\em before} $t=0$ whether the corresponding field 
mode was in fact in the Minkowski vacuum. She does that by using 
an inertial detector $D_1$, which must be inside her lab for a sufficient 
amount of time that she can verify the detector and 
the state of the relevant localized wavepacket. 

Bob uses a similar procedure by traveling on the ``mirror
trajectory'' (accelearating in the opposite direction) 
such
that he is detecting the other half of the Unruh entangled state.
\end{quote}

\begin{figure}\leavevmode 
\epsfxsize=8cm \epsfbox{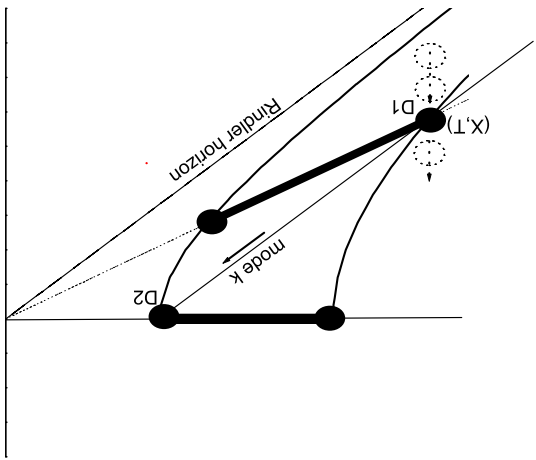} 
\caption{Upside-down spacetime diagram indicating coin flipping using the
Unruh effect.}
\end{figure}

The above protocol is slightly unconventional. However,
intuitively speaking, it is simply relying on the fact that what
Minkowski observers consider to be a separable vacuum state
$\left| 0\right\rangle_A \left| 0\right\rangle_B $, transforms to
the entangled state (\ref{ent}) for accelerating observers. The
purpose of the first detector measurement at time $t=0$ is to
ensure that the mode $k$ (see Fig.~1) which will determine the
coin flip outcome is in the correct Minkowski initial state. (A
cheating Bob could have his friend try and populate this mode
prior to it entering Alice's laboratory, for example).

Several other technical issues arise. Clearly $a$ must be chosen
to ensure that the probability of detector $D_2$ registering 0
photons is the same as that of registering one or more photons --
i.e. such that $\mu ^{2}=1/2$. We also wish this detector to be
sensitive to a localized, travelling Unruh wavepacket. A formal
quantization of Rindler space in terms of such modes can be found
in \cite{audretsch}, and from these results one can infer the
appropriate detector mode function responses required.

Another issue is that localized detectors necessarily will 
register ``dark counts'', with some (small) probability. This would affect
measurements done at small accelerations, but not at the impractically large
accelerations we are considering.

We conclude with a few observations. Firstly we must assume that
Alice's laboratory is large enough to contain both detectors at
the appropriate spacetime points. (Alice does not, however,
require other guarantees about the whereabouts of Bob). Also, the
further Alice and Bob are apart, the larger the acceleration has
to be. Thus there are nontrivial tradeoffs between the various
physical requirements of such a protocol.

\subsection{Other protocols}

Here we briefly consider other protocols of both types mentioned
previously:

\subsubsection{Dense coding}
Dense coding \cite{dense} is a protocol that allows one to send 2 classical bits of information
by sending one qubit, provided one prepared an entangled state in advance.
Here, however, at least the receiver would have to keep accelerating so as to
keep the entangled state, but in that case the receiver cannot receive anything
from the sender. Thus, it seems not possible to use the Unruh effect for dense coding.

\subsubsection{Quantum Key Distribution}
In Quantum Key Distribution Alice and Bob wish to share secret classical bits.
The way they can agree on classical bits is very much as in the coin flipping protocol. This time, though, they do trust each other, but not a possible eavesdropper Eve.
Just as in the Ekert protocol \cite{ekert},
they can check for Eve's existence by performing the appropriate Bell measurements. Indeed, it is well-known Bell inequalities are violated (to the maximum extent, in fact) in the vacuum of any quantum field theory \cite{werner}.
They do have to communicate classically, in order to check their measurement results,
but they can do that afterwards.

\subsubsection{Bit commitment}
As mentioned above, bit commitment \cite{bitc}
would require
even more than a trusted entangled state.
We suspect this protocol is not possible with the Unruh effect,
even though only classical
information is needed in the end.

\section{Summary}
Certain two-party
quantum communication protocols that require entanglement can be performed with just the vacuum.
Typically, protocols whose goal it is to produce a quantum state will not work in any useful way, but two-party protocols aimed at establishing purely classical information may work.
In particular, unconditionally secure coin flipping is possible, so is unconditionally secure
key distribution. On the other hand, teleportation between two accelerating observers is only possible in a weaker version.

\end{document}